\documentclass[prd,nofootinbib,showpacs,twocolumn]{revtex4}
\usepackage{color}
\usepackage{bm}
\usepackage{amsmath}
\usepackage{amssymb}
\usepackage{enumitem}
\usepackage{hyperref}
\usepackage{subfigure}
\usepackage{array}
\usepackage[english]{babel}
\usepackage{tensor}
\usepackage{comment}
\usepackage[normalem]{ulem}
\usepackage[dvipsnames]{xcolor}

\allowdisplaybreaks

\def\be{\begin{equation}}
\def\ee{\end{equation}}
\def\ba{\begin{eqnarray}}
\def\ea{\end{eqnarray}}

\def\f{\frac}

\usepackage{tikz}

\begin{document}

\title{Cosmological data favor  
Galileon ghost condensate over $\Lambda$CDM}

\author{Simone Peirone$^{1}$, 
Giampaolo Benevento$^{2,3}$, Noemi Frusciante$^{4}$, Shinji Tsujikawa$^{5}$\\ }

\affiliation{ 
\smallskip
$^{1}$Institute Lorentz, Leiden University, PO Box 9506, Leiden 2300 RA, The Netherlands
\smallskip\\
$^{2}$Dipartimento di Fisica e Astronomia ``G. Galilei'', Universit\`a degli Studi di Padova, via Marzolo 8, I-35131, Padova, Italy
\smallskip\\
$^{3}$INFN, Sezione di Padova, via Marzolo 8, I-35131, Padova, Italy
\smallskip\\
$^{4}$Instituto de Astrof\'isica e Ci\^encias do Espa\c{c}o, Faculdade de Ci\^encias da Universidade de Lisboa,  \\
 Edificio C8, Campo Grande, P-1749016, Lisboa, Portugal 
\smallskip \\
$^{5}$Department of Physics, Faculty of Science, Tokyo University of Science,
1-3, Kagurazaka, Shinjuku-ku, Tokyo 162-8601, Japan }

\begin{abstract}
We place observational constraints on the Galileon ghost condensate model, 
a dark energy proposal in cubic-order Horndeski theories consistent with the 
gravitational-wave event GW170817.
The model extends the covariant Galileon by taking 
an additional higher-order field derivative $X^2$ 
into account.
This allows for the dark energy equation of 
state $w_{\rm DE}$ to access the region $-2<w_{\rm DE}<-1$ without ghosts.
Indeed, this peculiar evolution of $w_{\rm DE}$ is favored over that of the cosmological constant $\Lambda$ 
from the joint data analysis 
of cosmic microwave background (CMB) radiation, baryonic acoustic oscillations (BAOs), 
supernovae type Ia (SNIa) and 
redshift-space distortions (RSDs). Furthermore, our model exhibits 
a better compatibility with the CMB data over the 
$\Lambda$-cold-dark-matter ($\Lambda$CDM) model by suppressing large-scale temperature anisotropies.  
The CMB temperature and polarization data lead to an estimation for today's Hubble parameter $H_0$ 
consistent with its direct measurements at  2$\sigma$.
We perform a model selection analysis by using several 
methods and find a statistically significant preference of 
the Galileon ghost condensate model over $\Lambda$CDM.
\end{abstract}


\date{\today}

\maketitle

\section{Introduction} 

The late-time cosmic acceleration has been firmly confirmed by several independent observations including 
SNIa \cite{SN1,SN2,Betoule}, 
CMB \cite{WMAP,Planck1,Planck2}, 
and BAOs \cite{Eisen,BAO1,BAO2}.
Although the cosmological constant $\Lambda$ is the 
simplest candidate for the source of this phenomenon, it is generally 
plagued by the problem of huge difference between 
the observed dark energy scale and the vacuum energy 
associated with particle physics \cite{Weinberg}. 
In the $\Lambda$CDM model, there have been also 
tensions for today's Hubble expansion rate 
$H_0$ constrained from the Planck CMB 
data \cite{Planck1} and 
its direct measurements at low redshifts \cite{Riess18}. 

In the presence of a scalar field $\phi$, the negative pressure 
arising from its potential or nonlinear kinetic energy 
can drive the cosmic acceleration. 
If we allow for derivative interactions and nonminimal 
couplings to gravity, Horndeski theories \cite{Horndeski} are the most general scalar-tensor theories with second-order equations of motion ensuring 
the absence of Ostrogradski instabilities \cite{Ho1,Ho2}. 
The gravitational-wave event GW170817 \cite{GW170817} together 
with its electromagnetic counterpart \cite{Goldstein} show 
that the speed of 
gravity $c_t$ is close to that of light with the relative difference 
$\sim 10^{-15}$. If we strictly demand that $c_t=1$, the Horndeski Lagrangian is 
of the form $L_{\rm H}=G_4(\phi)R+G_2(\phi,X)
+G_3(\phi, X) \square \phi$, 
where 
$R$ is the Ricci scalar, $G_4$ is a function of $\phi$, and 
$G_2, G_3$ depend on both $\phi$ and 
$X=\partial_\mu\phi \partial^\mu\phi$  \cite{GWcon1,GWcon2,GWcon3,GWcon4,GWcon5}.

Theories with the nonminimal coupling $G_4(\phi)R$ include 
$f(R)$ gravity and Brans-Dicke theories, but we have not yet found any observational signatures for supporting nonminimally coupled dark energy models over 
the cosmological constant.
The minimally coupled quintessence and k-essence with 
the Lagrangian $L=M_{\rm pl}^2 R/2+G_2(\phi,X)$, where $M_{\rm pl}$ 
is the reduced Planck mass, predicts $w_{\rm DE}>-1$ under the absence 
of ghosts, but there has been no significant observational 
evidence that these models are favored over $\Lambda$CDM.

The cubic-order Horndeski Lagrangian $G_3(\phi, X) \square \phi$ allows 
an interesting possibility for realizing $w_{\rm DE}<-1$ 
without ghosts.
In cubic Galileons with the Lagrangian 
$L=M_{\rm pl}^2 R/2+a_1 X+3a_3 X \square \phi$ \cite{Nicolis,Galileons}, where $a_1$ and $a_3$ are constants, there exists a tracker solution 
along which $w_{\rm DE}=-2$ during the matter era \cite{DT10}.
This behavior of $w_{\rm DE}$ is in tension with 
the joint data analysis of SNIa, CMB, and BAO \cite{NDT10}. 
The dominance of cubic Galileons as a dark energy density 
at low redshifts also leads to the enhancement of perturbations 
incompatible with measurements of the cosmic growth history \cite{Renk,Peirone}.

The above problems of Galileons are alleviated by taking a scalar potential $V(\phi)$ \cite{Ali,KTD15} or 
a nonlinear term of $X$ in $G_2(\phi,X)$ 
into account \cite{Kase18}. 
In particular, the latter model can lead to $w_{\rm DE}$ 
in the range $-2<w_{\rm DE}<-1$.
Moreover, the Galileon is not necessarily the main source for 
late-time cosmic acceleration in this case, so it should be compatible with  
cosmic growth measurements. In this letter, we show that 
the cubic Galileon model with a nonlinear term in $X$ 
exhibits a novel feature of being observationally 
favored over $\Lambda$CDM.

\section{Model}

We study the Galileon ghost condensate (GGC)
model given by the action
\be
{\cal S}=\int {\rm d}^4 x \sqrt{-g} \left[ \frac{M_{\rm pl}^2}{2}R
+a_1 X+a_2 X^2+3a_3X \square \phi \right]+{\cal S}_M,
\label{action}
\ee
where $a_{1,2,3}$ are constants.
For the matter action ${\cal S}_M$, we consider perfect fluids 
minimally coupled to gravity. 
The existence of term $a_2X^2$ leads to the modified evolution 
of $w_{\rm DE}$ and different cosmic growth history 
compared to those of the cubic Galileon 
(which corresponds to $a_2=0$). 
The ghost condensate model \cite{Arkani} can be recovered 
by taking the limit $a_3 \to 0$ in Eq.~(\ref{action}). 

On the flat Friedmann-Lemaitre-Robertson-Walker (FLRW) 
background given by the line element 
${\rm d} s^2=-{\rm d}t^2+a^2(t)\delta_{ij}{\rm d}x^i {\rm d}x^j$, 
we consider nonrelativistic matter (density $\rho_m$ with vanishing pressure)
and radiation  (density $\rho_r$ and pressure $P_r=\rho_r/3$)
for the action ${\cal S}_M$. 
To discuss the background cosmological dynamics, 
it is convenient to introduce the dimensionless variables
\be
x_1=-\f{a_1\dot{\phi}^2}{3M_{\rm pl}^2 H^2}\,,\quad x_2=\f{a_2\dot{\phi}^4}{M_{\rm pl}^2 H^2}\,, \quad x_3=\f{6a_3\dot{\phi}^3}{M_{\rm pl}^2H}\, ,\label{dimensionless_functions}
\ee
where $H=\dot{a}/a$, and a dot represents the derivative with respect to 
the cosmic time $t$. Then, the Friedmann equation can be expressed in the form $
\Omega_m+\Omega_r+
\Omega_{\rm DE}=1$\,
where $\Omega_m=\rho_m/(3M_{\rm pl}^2 H^2)$, 
$\Omega_r=\rho_r/(3M_{\rm pl}^2 H^2)$, and 
\be\label{consteq}
\Omega_{\rm DE}=x_1+x_2+x_3\,.
\ee
The variables $x_1$, $x_2$, $x_3$, and $\Omega_r$ 
correspond to density parameters associated with the Lagrangians 
$a_1 X$, $a_2 X^2$, $3a_3 X \square \phi$, and radiation, respectively. Equation \eqref{consteq} evaluated 
today allows us to eliminate one free parameter, leaving the model with  two extra parameters compared to  $\Lambda$CDM.

The dynamical system can be expressed in the form 
\ba
&&x_1'=2x_1(\epsilon_\phi-h)\,,\qquad 
x_2'=2x_2(2\epsilon_\phi-h)\,, \nonumber \\
&&x_3'=x_3(3\epsilon_\phi-h)\,,\qquad
\Omega_r'=-2\Omega_r(2+h)\,,
\label{system}
\ea
where $\epsilon_{\phi}=\ddot{\phi}/(H \dot{\phi})$, 
$h=\dot{H}/H^2$, and a prime represents a derivative with 
respect to ${\cal N}=\ln a$. 
The explicit expressions of $\epsilon_{\phi}$ and $h$ 
are given in Eqs.~(4.16) and (4.17) of Ref.~\cite{Kase18} 
(with $x_4=0$).
The dark energy equation of state is 
\be
w_{\rm DE}=\frac{3 x_1+x_2-\epsilon_{\phi} x_3}
{3(x_1+x_2+x_3)}\,.
\label{wde}
\ee

On the future de Sitter fixed point we have  
$\Omega_{\rm DE}=1$, and $w_{\rm DE}=-1$ 
with $\epsilon_{\phi}=0$, so 
there are two relations 
$x_1^{\rm dS}=-2+x_3^{\rm dS}/2$ and 
$x_2^{\rm dS}=3-3x_3^{\rm dS}/2$. 
Even though $x_1^{\rm dS}$ is negative 
for $x_3^{\rm dS} \ll 1$, the ghost can be avoided by 
the positive $x_2^{\rm dS}$ term.

If the condition $x_3 \gg \{ |x_1|, x_2 \}$ is satisfied 
in the early cosmological epoch, we have
$w_{\rm DE} \simeq -\epsilon_{\phi}/3 \simeq 1/4-\Omega_r/12>0$. 
On the other hand, in the limit $x_2 \to 0$, there exists a tracker solution 
satisfying the relation $x_3=-2 x_1$ (or equivalently, 
$\epsilon_{\phi}=-h$) \cite{DT10,Kase18}. 
In this case, Eq.~(\ref{wde}) reduces to 
$w_{\rm DE}=-1+2h/3$ and hence $w_{\rm DE} \simeq -2$ 
during the matter era. The existence of positive $x_2$ 
can lead to $w_{\rm DE}$ larger than $-2$, so the approach 
to the tracker is prevented by the term $a_2 X^2$. 
Indeed, after $x_2$ catches up with $x_3$, the solutions 
tend to approach the de Sitter attractor with $x_3$ 
subdominant to $|x_1|$ and $x_2$ at low redshifts \cite{Kase18}. 
In this way, the background dynamics temporally entering 
the region $-2<w_{\rm DE}<-1$ can be realized by the 
model (\ref{action}) with $a_2 \neq 0$.

\section{Cosmological perturbations}

For the GGC model (\ref{action}), 
the propagation of tensor perturbations is the same as 
that in General Relativity (GR).  
As for scalar perturbations, we consider the
perturbed line element on the flat FLRW background:
\be
{\rm d}s^2=-\left( 1+2\Psi \right) {\rm d}t^2
+a^2(t) \left(1-2\Phi \right) \delta_{ij} {\rm d}x^i 
{\rm d}x^j\,,
\ee
where $\Psi$ and $\Phi$ are gravitational potentials.
In Fourier space with the coming wavenumber $k$, 
we relate $\Psi$ and $\Psi+\Phi$ with the total 
matter density perturbation 
$\rho \Delta=\sum_{i} \rho_i \Delta_i$ (where $i=m,r,...$),  
as \cite{Amen07,Ber08,Pogo10}
\ba 
\label{mudef}
&&-k^2\Psi=4\pi G_{\rm N} a^2\mu(a,k)\rho\Delta\,, \\
&&-k^2(\Psi+\Phi)=8\pi G_{\rm N} 
a^2\Sigma(a,k)\rho\Delta\,,
\ea
where $G_{\rm N}=(8\pi M_{\rm pl}^2)^{-1}$ is the Newtonian gravitational constant. 
The dimensionless quantities $\mu$ and $\Sigma$ characterize 
the effective gravitational couplings felt by matter and 
light, respectively. 
Applying the quasi-static approximation \cite{Pola00,DKT} for perturbations deep inside the Hubble radius to the 
model (\ref{action}), it follows that \cite{Kase18}
\be
\mu=\Sigma=1+\frac{x_3^2}{Q_s c_s^2 (2-x_3)^2}\,,
\ee
where
\ba
Q_s &=& \frac{3(4x_1+8x_2+4x_3+x_3^2)}
{(2-x_3)^2}\,,\\
c_s^2 &=& \frac{2(1+3\epsilon_{\phi})x_3-x_3^2
-4h-6\Omega_m-8\Omega_r}
{3(4x_1+8x_2+4x_3+x_3^2)}\,.
\ea
To avoid ghosts and Laplacian instabilities, 
we require that $Q_s>0$ 
and $c_s^2>0$. Then, for $x_3 \neq 0$, 
$\mu$ and $\Sigma$ are larger than 1, so both 
$\Psi$ and $\Psi+\Phi$ are enhanced compared 
to those in GR.
Since $\mu=\Sigma$, there is no gravitational 
slip ($\Psi=\Phi$). For the sub-horizon perturbations, 
the matter density contrast $\Delta$ approximately obeys
\be
\ddot{\Delta}+2H \dot{\Delta}
-4\pi G_{\rm N} \mu \rho\Delta=0\,,
\ee
so the cosmic growth rate is larger than that in GR. In the likelihood analysis, we solve full perturbation equations without resorting to the quasi-static approximation.  

\section{Methodology of cosmological probes}

To confront the GGC model with observations, we use the Planck 2015 data of CMB temperature anisotropies 
and polarizations \citep{Planck1,Planck2}. For the Planck likelihood, we also 
vary the nuisance parameters exploited to model foregrounds as well as instrumental and beam uncertainties. 
We consider the former dataset in combination with data from the CMB lensing reconstruction~\cite{Ade:2015zua}, 
to which we refer as ``Planck+Lensing".
We include the BAO data from the 6dF galaxy survey~\citep{BAO1} 
and the SDSS DR7 main galaxy sample~\citep{BAO2}. 
Furthermore, we employ the combined BAO and RSD
data from the SDSS DR12
consensus release \cite{Alam}, together with the JLA SNIa 
sample~\cite{Betoule}. The latter dataset is called ``PBRS''. 

We modify the public available Einstein-Boltzmann code EFTCAMB~\cite{Hu:2013twa,Raveri:2014cka}
by  implementing a background solver and mapping relations for the chosen model following 
the prescription in Refs.~\cite{Gubitosi:2012hu,Bloomfield:2012ff,Gleyzes:2013ooa,Frusciante:2016xoj}.  
The built-in stability module  allows us to identify the viable parameter space by imposing 
the two stability conditions $Q_s>0$ and $c_s^2>0$.
These results will be used to set priors for the data analysis.
We impose flat priors on the initial values of 
two model  parameters: 
$x_1^{(i)} \in [-10, 10] \times 10^{-16}$, 
$x_3^{(i)} \in [-10, 10]\times 10^{-9}$ 
at the redshift $z=10^{5}$.
We performed a test simulation in which the prior ranges
are increased by one order of magnitude and 
found no difference for the likelihood results. 

\section{Observational constraints}

\begin{figure}
\includegraphics[height=2.9in,width=3.4in]{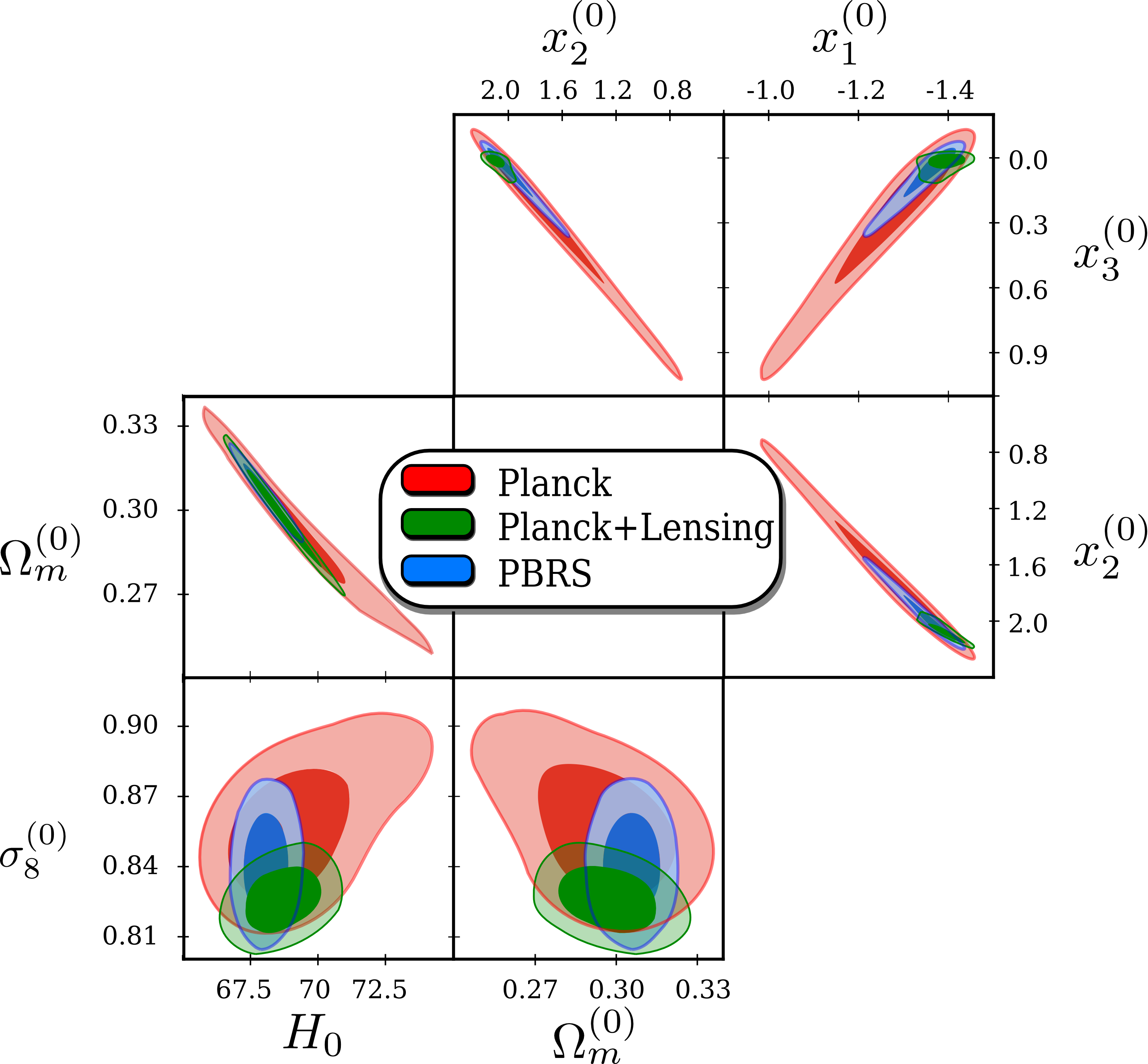}
\caption{Joint marginalised constraints 
(68\,\% and 95\,\% CLs) on six model parameters
$x_1^{(0)}, x_2^{(0)}, x_3^{(0)}, H_0, 
\sigma_8^{(0)}, \Omega_m^{(0)}$ 
obtained with the Planck, Planck+Lensing, 
and PBRS datasets. 
\label{fig:contours} 
}
\end{figure}

In Tables \ref{tab:best_fit_model} and \ref{tab:best_fit_cosmo}, we show 
today's values $x_1^{(0)}, x_2^{(0)}, x_3^{(0)}$ and 
$H_0$, $\sigma_8^{(0)}$, $\Omega_m^{(0)}$
constrained from the Planck and PBRS datasets, 
together with bounds on the latter three parameters 
in $\Lambda$CDM.
In Fig.~\ref{fig:contours}, we also plot two-dimensional observational bounds 
on six parameters by including the Planck+Lensing data as well.
In GGC, the Planck data alone lead to higher values of $H_0$ than that in $\Lambda$CDM. 
The former model is consistent with the  Riess {\it et al.} bound $H_0=73.48 \pm 1.66$~km~s$^{-1}$ Mpc$^{-1}$ 
derived by direct measurements of $H_0$ using Cepheids \cite{Riess18}. 
With the PBRS and CMB lensing datasets,  we find that the bounds on 
$H_0$, $\sigma_8^{(0)}$ and $\Omega_m^{(0)}$ 
are compatible between GGC and $\Lambda$CDM. 
We do not include the data of direct measurements of $H_0$ and 
weak lensing, 
as they can be affected by the statistical analysis \cite{Efstathiou} 
and nonlinear perturbation dynamics  \cite{Hildebrandt:2016iqg}, respectively. 

\begin{table}[t!]
\centering
\begin{tabular}{|c|c|c|}
\hline
Parameter  & Planck & PBRS \\
\hline
\hline 
 &  &   \\[-8pt]
$x_1^{(0)}$ & $ -1.27^{+0.22}_{-0.15} \,(-1.26) $ & $ -1.35^{+0.1}_{-0.07} \,(-1.27)  $  \\[-8pt]
 &  &   \\
\hline
 &  &   \\[-8pt]
$x_2^{(0)}$ & $ 1.70^{+0.45}_{-0.73}\,(1.64)  $ & $ 1.95^{+0.18}_{-0.31}\, (1.74)  $  \\[-8pt]
 &  &   \\
\hline
 &  &   \\[-8pt]
$x_3^{(0)}$ & $0.28^{+0.5}_{-0.3}\,(0.34)  $ & $ 0.09^{+0.2}_{-0.1} \,(0.23) $  \\[-8pt]
 &  &   \\
\hline
\end{tabular}
\caption{Marginalized values of the model parameters $x_1^{(0)}, x_2^{(0)}, x_3^{(0)}$ and 
their $95\%$ CL bounds, obtained by 
Planck and PBRS datasets. In parenthesis we show maximum likelihood  values.}
\label{tab:best_fit_model}
\end{table}
\begin{table}[t!]
\centering
\begin{tabular}{|c|c|c|c|}
\hline
Parameter & Case & Planck & PBRS \\
\hline
\hline
 & GGC & $ 69.3^{+3.6}_{-3.0} \,(70) $ & $ 68.1 \pm 1.1 \,(68.4) $  \\[-6pt]
$H_0$ &  &   &    \\[-6pt]
& $\Lambda$CDM & $  67.9 \pm 2.0\,(67.6) $ & $ 68\pm 1 \,(68) $  \\
\hline
&  GGC & $0.86 \pm0.04 \,(0.87)$ & $0.84 \pm 0.03$ \,(0.85) \\[-6pt]
$\sigma_8^{(0)}$  &    &   &   \\[-6pt]
&  $\Lambda$CDM & $0.841 \pm 0.03 \,(0.83)$ & $ 0.84 \pm 0.03\,(0.84) $ \\
\hline
&  GGC & $0.30\pm 0.04 \,(0.28)$ & $ 0.305 \pm 0.01$ (0.30)  \\[-6pt]
$\Omega_m^{(0)}$  &   &   &   \\[-6pt]
&  $\Lambda$CDM & $ 0.30 \pm 0.03\,(0.31) $ & $ 0.31 \pm 0.01\,(0.31) $ \\
\hline
\end{tabular}
\caption{Marginalized values of 
$H_0$, $\sigma_8^{(0)}$, $\Omega_m^{(0)}$
and their $95\%$ CL bounds. }
\label{tab:best_fit_cosmo}
\end{table}

\begin{figure}
\includegraphics[width=.49\textwidth]{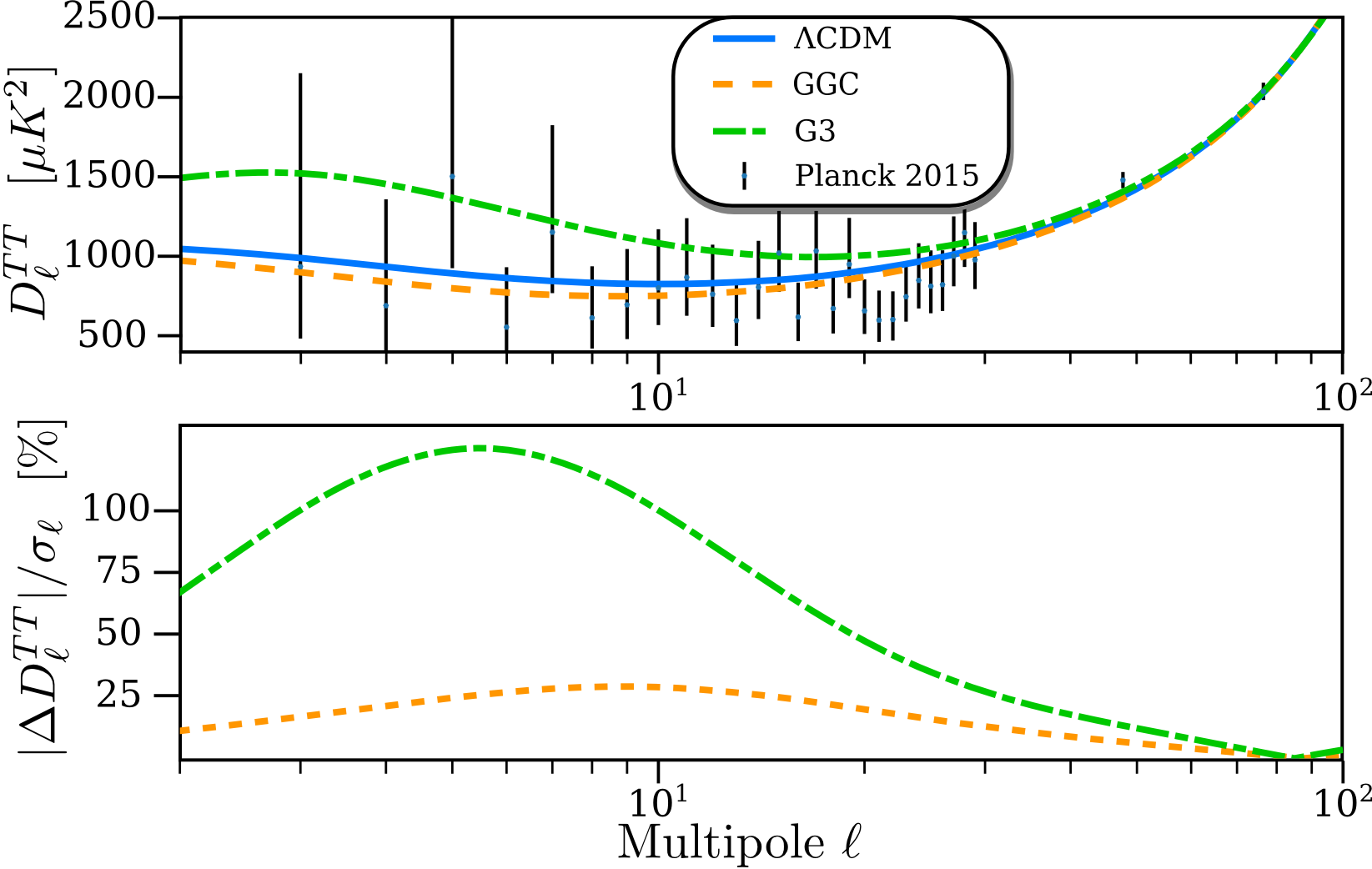}
\caption{\textit{Top panel}: 
Best-fit CMB temperature-temperature (TT) power spectra
$D_\ell^{\rm TT} = \ell(\ell+1)/2 \pi C_\ell^{\rm TT}$ at low multipoles $\ell$
for $\Lambda$CDM, GGC, and G3 (cubic Galileons), as obtained in the analysis 
of the Planck dataset. 
The best-fit values for G3 are taken from Ref.~\cite{Peirone}.
For comparison, we plot the data points from Planck 2015. 
\textit{Bottom panel}: 
Relative difference of the best-fit TT power spectra, 
in units of cosmic variance 
$\sigma_\ell = \sqrt{2/(2 \ell+1)} C_\ell^{\Lambda {\rm CDM}}$.
\label{fig:bestfitTT} 
}
\end{figure}

The values of $x_1^{(0)}$ and $x_2^{(0)}$ constrained 
from the data are of order 1, with $x_1^{(0)}<0$ 
and $x_2^{(0)}>0$. 
We find the upper limit 
$x_3^{(0)}<0.118$ (68\,\% CL) from the PBRS data.
This bound mostly arises from the fact that the 
dominance of $x_3$
over $x_2$ at low redshifts leads to the 
enhanced Integrated Sachs-Wolfe (ISW) effect on 
CMB temperature anisotropies. 
The most stringent constraints on model parameters are obtained with 
the Planck+Lensing datasets.
In Fig.~\ref{fig:bestfitTT}, we plot the CMB TT power 
spectra for GGC as well as for 
$\Lambda$CDM and cubic Galileons (G3), given by the best-fit to the Planck data.
The G3 model corresponds to $x_2=0$, so that the 
Galileon density is the main source for 
cosmic acceleration. 
In this case, the TT power spectrum for the multipoles 
$l<{\cal O}(10)$ is strongly enhanced relative to 
$\Lambda$CDM and this behavior is disfavored from 
the Planck data \cite{Peirone}.

\begin{figure}
\includegraphics[height=2.3in,width=3.4in]{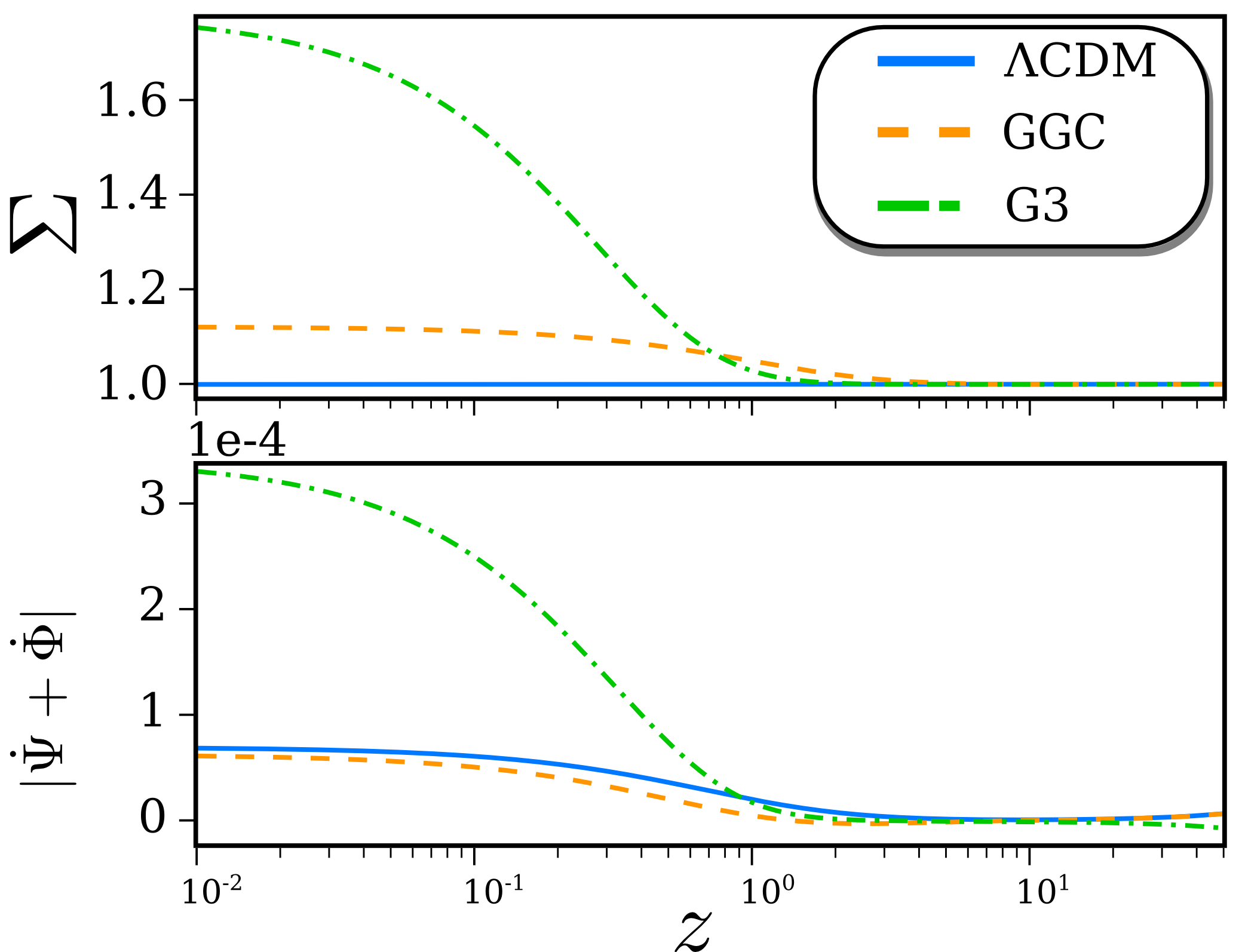}
\caption{Best-fit evolution of $\Sigma$ (top) and 
$|\dot{\Psi}+\dot{\Phi}|$ (bottom) versus 
$z$ at $k = 0.01$ Mpc$^{-1}$ for  $\Lambda$CDM, GGC, and G3 
derived with the PBRS dataset. 
\label{fig:bestfitsig} 
}
\end{figure}

\begin{figure}
\includegraphics[height=1.8in,width=3.3in]{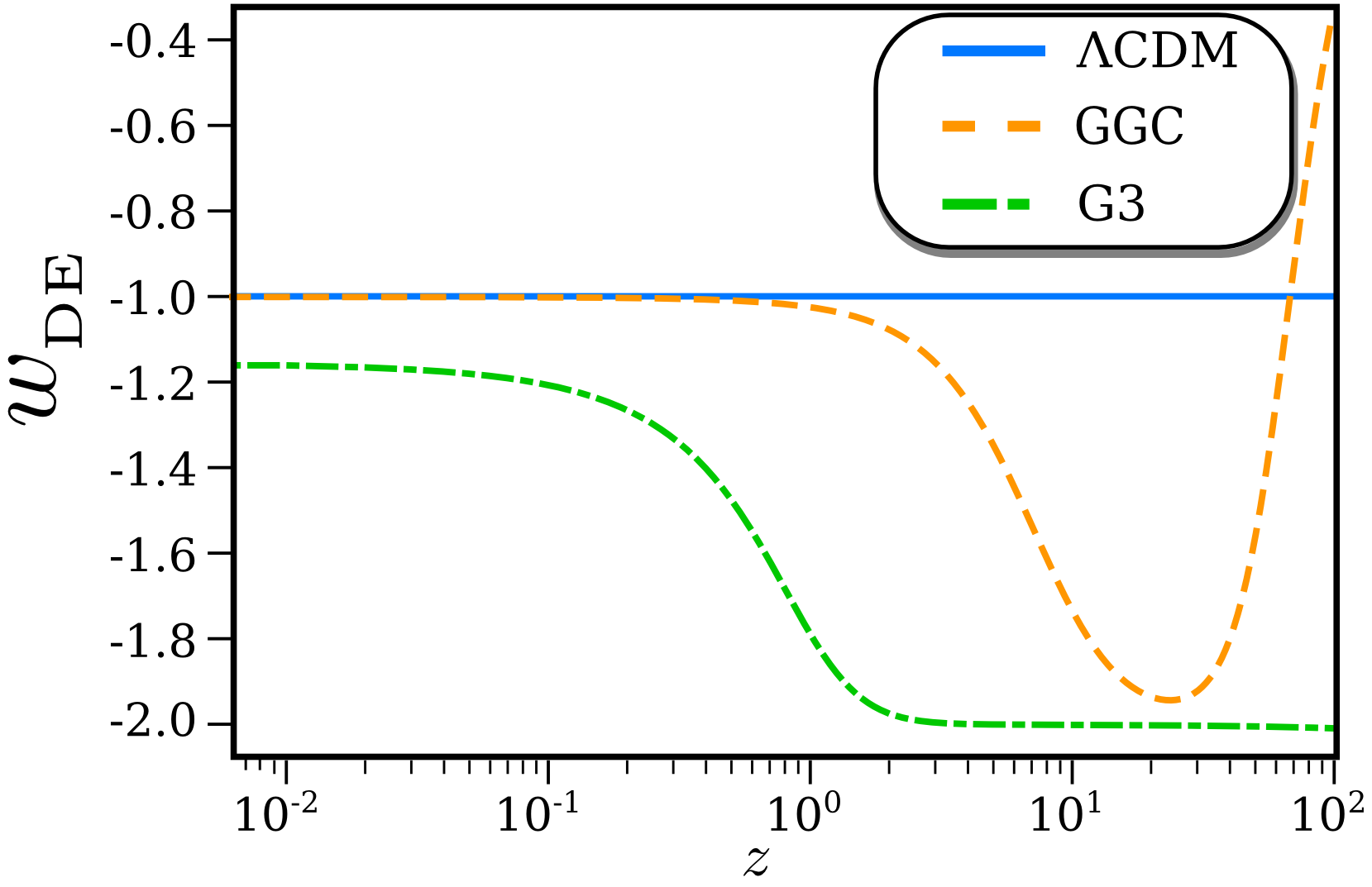}
\caption{Best-fit evolution of $w_{\rm DE}$ versus 
$z$ for $\Lambda$CDM, GGC, and G3 
derived with the PBRS dataset.
\label{fig:bestfitw} 
}
\end{figure}

In GGC, the $a_2 X^2$ term in (\ref{action}) can 
avoid the dominance of $x_3$ over $x_2$ around today.
Even if $x_3^{(0)} \ll x_2^{(0)}$, the cubic Galileon 
gives rise to an interesting contribution to the
CMB TT spectrum. As we see in Fig.~\ref{fig:bestfitTT}, 
the best-fit GGC model is in better agreement with the Planck data relative to 
$\Lambda$CDM by suppressing large-scale ISW tails. 
Taking the limit $x_3^{(0)} \to 0$, 
the TT spectrum approaches the one in $\Lambda$CDM.
The TT spectrum of G3 in Fig.~\ref{fig:bestfitTT} can be recovered by taking 
the limit $x_3^{(0)} \gg x_2^{(0)}$.

In Fig.~\ref{fig:bestfitsig}, we depict the evolution of $\Sigma$ and 
$|\dot{\Psi}+\dot{\Phi}|$ for GGC, G3 and $\Lambda$CDM, obtained from the PBRS best-fit. 
In G3, the large growth of $\Sigma$ from 1  
leads to the enhanced ISW effect on CMB anisotropies determined 
by the variation of $\Psi+\Phi$ at low redshifts. 
For the best-fit GGC, the deviation of $\Sigma$ from 1 
is less significant, with $\dot{\Psi}+\dot{\Phi}$ closer to 0.
In the latter case, the TT spectrum is suppressed with respect to 
$\Lambda$CDM. This is why 
the intermediate value of $x_3^{(0)}$ around 0.1 
with $x_2^{(0)}={\cal O}(1)$ exhibits the better 
compatibility with the CMB data relative to $\Lambda$CDM.

As we see in Fig.~\ref{fig:bestfitw}, the best-fit GGC 
corresponds to the evolution of  $w_{\rm DE}$ approaching 
the asymptotic value $-1$ from the region 
$-2<w_{\rm DE}<-1$. This overcomes the problem of 
G3 in which the $w_{\rm DE}=-2$ behavior during the 
matter era is inconsistent with the CMB+BAO+SNIa 
data \cite{NDT10}.
This nice feature of $w_{\rm DE}$ in GGC again
comes from the combined effect of $x_2$ and $x_3$.

\section{Model selection}

The GGC model has two extra parameters with respect to $\Lambda$CDM, 
to allow for a better fit to the data.
In order to determine whether GGC is favored over 
$\Lambda$CDM, 
we make use of the 
Deviance Information Criterion (DIC)~\cite{RSSB}:
\be
{\rm DIC} = \chi^2_{\rm eff} (\hat{\theta}) + 2 p_{\rm D}\,,
\ee
where $\chi^2_{\rm eff} (\hat{\theta}) = -2 \ln \mathcal{L}(\hat{\theta})$ with $\hat \theta$ being 
parameters maximizing the likelihood function 
$\mathcal{L}$, and 
$p_{\rm D} = \bar{ \chi}^2_{\rm eff} ({\theta}) -  \chi^2_{\rm eff} (\hat{\theta})$. 
Here, the bar denotes an average over the posterior distribution. We observe that the DIC accounts for both the goodness of fit, 
$\chi^2_{\rm eff} (\hat{\theta})$, and for the Bayesian complexity of the model, 
$ p_{\rm D}$, which disfavors more complex models. 
For the purpose of model comparisons, 
we compute
\be
\Delta {\rm DIC} =  {\rm DIC} _{\rm GGC} - {\rm DIC} _{\Lambda{\rm CDM}},
\ee
from which we infer that a negative (positive) $\Delta {\rm DIC}$ 
would support GGC ($\Lambda$CDM).  

We also consider the Bayesian evidence factor ($\log_{10}B$) 
along the line of Refs.~\cite{Heavens:2017afc,DeBernardis:2009di} to quantify the support for GGC over $\Lambda$CDM.  
A positive value of $\Delta \log_{10}B $ indicates a statistical preference for the extended model and a
 strong preference is defined for $\Delta \log_{10}B > 2$.

\begin{table}[h]
\centering
\begin{tabular}{|c|c|c|c|c|}
\hline
Dataset &  $\Delta \chi_{\rm eff} ^2$ & $\Delta {\rm DIC} $ & $\Delta \log_{10} B $\\
\hline
Planck  &$-4.8$& $-2.5$ & 4.4\\
\hline
PBRS &  $-2.8$ & $-0.6$ & 5.1 \\
\hline
Planck+Lensing &$-0.9$ &0.80 & 1.6\\
\hline
\end{tabular}
\caption{Model comparisons through the obtained values of $\Delta \chi_{\rm eff} ^2$, 
$\Delta {\rm DIC}$ and $\Delta \log_{10} B$ using $\Lambda$CDM as reference.}
\label{tab:model_selection}
\end{table}
 
In Table~\ref{tab:model_selection}, we list the values of 
$\Delta \chi_{\rm eff} ^2$, $\Delta {\rm DIC}$ 
and $\Delta \log_{10}B$ computed with respect to 
$\Lambda$CDM for each dataset considered in this analysis. For Planck and  PBRS
both $\Delta {\rm DIC}$ and $\Delta \log_{10}B$ exhibit significant preferences for GGC over $\Lambda$CDM. 
This suggests that not only the CMB data but also 
the combination of BAO, SNIa, RSD datasets 
favors the cosmological dynamics of GGC 
like the best-fit case shown in Figs.~\ref{fig:bestfitsig} 
and \ref{fig:bestfitw}. 
With the Planck+Lensing data the $\chi_{\rm eff} ^2$ and Bayesian factor 
exhibit slight preferences for GGC, while 
the DIC mildly favours $\Lambda$CDM.
The model selection analysis with the CMB lensing data does not give a definite conclusion for the preference of models. We note that, among the likelihoods used in our analysis, the CMB lensing alone assumes $\Lambda$CDM as a fiducial model~\cite{Ade:2015zua}. This might source a  bias towards the latter.

\section{Conclusion}

We have shown that, according to the two 
information criteria, GGC is significantly
favoured over $\Lambda$CDM with the PBRS datasets. 
This property holds even with two additional model parameters 
than those in $\Lambda$CDM. 
According to our knowledge, there are no other scalar-tensor dark energy 
models proposed so far showing such novel properties.
This surprising result is attributed to the properties that, 
for $x_3^{(0)} \ll x_2^{(0)}={\cal O}(1)$,  
(i) suppressed ISW tails relative to $\Lambda$CDM 
can be generated, and 
(ii) $w_{\rm DE}$ can be in the region 
$-2<w_{\rm DE}<-1$ at low redshifts. 
The GGC model deserves for being tested further in future 
observations of WL, ISW-galaxy cross-correlations, 
and gravitational waves.

\section*{Acknowledgments}

We thank N.~Bartolo, A.~De Felice, R.~Kase, 
M.~Liguori, M.~Martinelli, S.~Nakamura, 
M.~Raveri and A.~Silvestri  for useful discussions.
SP acknowledges support from the NWO and the Dutch Ministry of Education, Culture and Science (OCW), and also from the D-ITP consortium, a program of the NWO that is funded by the OCW.
GB acknowledges financial support from Fondazione Ing.~Aldo~Gini. 
The research of NF is supported by Funda\c{c}\~{a}o para a  Ci\^{e}ncia e a Tecnologia (FCT) through national funds  (UID/FIS/04434/2013), by FEDER through COMPETE2020  (POCI-01-0145-FEDER-007672) and by FCT project ``DarkRipple -- Spacetime ripples in the dark gravitational Universe" with ref.~number PTDC/FIS-OUT/29048/2017.
SP, GB and NF acknowledge the COST Action  (CANTATA/CA15117), supported by COST (European Cooperation in  Science and Technology).
ST is supported by the
Grant-in-Aid for Scientific Research Fund of the JSPS No.~19K03854 and
MEXT KAKENHI Grant-in-Aid for Scientific Research on Innovative Areas
``Cosmic Acceleration'' (No.\,15H05890).

\end{document}